\documentclass[preprint,showpacs,preprintnumbers,amsmath,amssymb,nofootinbib]{revtex4}
\usepackage[dvipdfmx]{graphicx}
\usepackage{amsmath,amssymb}
\usepackage{bm}
\usepackage{color} 

\def\s{\sigma}

\def\up{\uparrow}
\def\dwn{\downarrow}

\def\lesssim{\ \raise.3ex\hbox{$<$}\kern-0.8em\lower.7ex\hbox{$\sim$}\ }
\def\gesim{\ \raise.3ex\hbox{$>$}\kern-0.8em\lower.7ex\hbox{$\sim$}\ }

\newcommand \beq{\begin{eqnarray}}
\newcommand \eeq{\end{eqnarray}}

\begin{document}
\title{Superfluid Phase Transitions and Effects of Thermal Pairing Fluctuations in Asymmetric Nuclear Matter}
\author{Hiroyuki Tajima$^1$, Tetsuo Hatsuda$^{2,1}$, Pieter van Wyk,$^3$ and Yoji Ohashi$^3$}
\affiliation{$^1$Quantum Hadron Physics Laboratory, RIKEN Nishina Center,
  Wako, Saitama 351-0198, Japan}
\affiliation{$^2$Interdisciplinary Theoretical and Mathematical Sciences Program (iTHEMS), RIKEN, Wako,
  Saitama 351-0198, Japan}
\affiliation{$^3$Department of Physics, Keio University, Hiyoshi, Kohoku-ku, Yokohama, 223-8522, Japan}
\begin{abstract}
We investigate superfluid phase transitions of asymmetric nuclear matter
 at finite temperature ($T$) and  density ($\rho$)
  with a  low proton fraction ($Y_{\rm p} \le 0.2$) which is relevant to the inner crust and outer core of neutron stars.
A strong-coupling theory developed  for two-component atomic Fermi gases is generalized to
 the  four-component case and is applied to the system of  spin-$1/2$  neutrons and protons.
 The empirical phase shifts of 
  neutron-neutron (nn), proton-proton (pp) and neutron-proton (np) interactions  up to $k = 2$ ${\rm fm}^{-1}$ 
   are described  by multi-rank separable potentials.
 We show that (i) the critical temperature of the neutron superfluidity $T_{\rm c}^{\rm nn}$ at $Y_{\rm p}=0$
agrees well with Monte Carlo data at low densities and takes a maximum value
$T_{\rm c}^{\rm nn}=1.68$ MeV at $\rho/\rho_0 = 0.14$ with $\rho_0=0.17$ fm$^{-3}$, (ii)
 the critical temperature of the proton superconductivity
$T_{\rm c}^{\rm pp}$ for $Y_{\rm p} \le 0.2$ is substantially 
suppressed at low densities due to np-pairing fluctuations and starts to dominate over
  $T_{\rm c}^{\rm nn}$ only  above $\rho/\rho_0 = 0.70$ $(0.77)$ for $Y_p =0.1$ $(0.2)$,
and (iii) the deuteron condensation temperature $T_{\rm c}^{\rm d}$
is suppressed at $Y_{\rm p}\le 0.2$ due to the large mismatch of the two Fermi surfaces.
\end{abstract}
\maketitle
\section{Introduction}
\label{sec1}

The superfluidity  in strongly interacting Fermi systems has attracted much attention both theoretically and experimentally.
For reviews, we refer to
Refs.~\cite{Takatsuka,Dean} in nuclear physics,
Refs.~\cite{Oertel,Page3,Baym:2017whm} in astrophysics, as well as Refs.~\cite{Carlson:2012mh,Gandolfi,Horikoshi,HorikoshiE,Strinati}
 in condensed matter physics. 
   It has been also recognized that the dilute neutron matter and  two-component ultracold atomic fermions near the unitarity 
  have close similarity to each other,
  due to the strong pairing interactions associated with the large negative neutron-neutron scattering length $a_{\rm s}=-18.5$ fm and relatively small effective range $r_{\rm eff}=2.7$ fm (see Refs.~\cite{Carlson:2012mh,Gandolfi,Horikoshi,HorikoshiE,Strinati} and references therein).
In the latter atomic system, the pairing interaction can be described by a zero-range potential with a large scattering length~\cite{Chin}.
  In strongly interacting systems, such as neutron matter and unitary Fermi gases, 
  effects of pairing fluctuations near the superfluid phase transition are particularly important.
 Such effects have extensively been studied in cold Fermi gas physics through the observations of various quantities, such as 
 single-particle excitation spectrum, specific heat, superfluid phase transition temperature~($T_{\rm c}$), shear viscosity, and spin susceptibility~\cite{Strinati,Mueller,Jensen}.
 Three of the present authors have recently shown~\cite{Pieter1} that a strong coupling 
theory, being based on the one developed by Nozi\`{e}res and Schmitt-Rink (NSR)~\cite{Nozieres}
   can provide a unified description of neutron matter and an ultracold Fermi gas in the unitary regime.
This indicates that the latter atomic gas system can be
  used as a quantum simulator for neutron star interiors at subnuclear densities.

There are, however, some issues  to be overcome for better understanding of  the physics of 
 neutron star interiors:
Besides neutrons, one should also include a non-zero fraction $Y_{\rm p}$ of protons.
To deal with this, one needs to extend strong-coupling theories developed for two-component atomic Fermi gases to the four-component case
involving spin and isospin degrees of freedom.
  In such a system, not only a neutron-neutron (nn) interaction but also a 
   proton-proton (pp) interaction, as well as a neutron-proton (np) interaction, work.
     In particular, the np interaction in the 
    deuteron channel is stronger than the other interactions, so that it may affect the onset of proton
    superconductivity.  Furthermore, 
     the short-range repulsion of the nuclear force is important
     to describe the pairing phenomena around the nuclear matter density. 
          In this paper, we will consider all these points and study
     the critical temperature of the superfluid phase transitions in asymmetric nuclear matter
      around the nuclear saturation density $\rho_0=0.17$ fm$^{-3}$, by including the nn, pp and np pairng fluctuations.

This paper is organized as follows.
In Sec. \ref{sec2}, we present our model for asymmetric nuclear matter, as well as details of our strong coupling scheme.
In Sec. \ref{sec3}, we show our numerical results for the critical temperatures associated with the nn, pp and np pairings as functions of nucleon density and proton fraction.
In this paper, we set $\hbar=k_{\rm B}=1$, and the system volume is taken to be unity, for simplicity.

\section{Formalism}
\label{sec2}
\subsection{Effective Hamiltonian}
We introduce  the pair operator
 $S_m$  ($T_\ell$)
  in the spin-singlet--isospin-triplet (spin-triplet--isospin-singlet) channel with the relative momentum~$\bm{k}$ and the center of mass   momentum~$\bm{q}$:
\beq
S_m (\bm{k},\bm{q}) &=& \sum_{\lambda, \rho} \sum_{i,j} 
\Bigl\langle \frac{1}{2} \frac{1}{2} \lambda \rho \Bigr|\Bigl.00 \Bigr\rangle 
\Bigl\langle \frac{1}{2} \frac{1}{2} ij \Bigr|\Bigl. 1 m \Bigr\rangle 
 c_{\bm{-k+q}/2,\lambda,i}  c_{\bm{k+q}/2,\rho,j} \\
T_\ell (\bm{k},\bm{q}) &=& \sum_{\lambda, \rho} \sum_{i,j} 
\Bigl\langle \frac{1}{2} \frac{1}{2} \lambda \rho \Bigr|\Bigl.1 \ell \Bigr\rangle 
\Bigr\langle \frac{1}{2} \frac{1}{2} ij \Bigr|\Bigl. 00 \Bigr\rangle 
 c_{\bm{-k+q}/2,\lambda,i}  c_{\bm{k+q}/2,\rho,j} 
 \eeq
Here  $ c_{\bm{k},\lambda,i}$  is the fermion annihilation operator with momentum $\bm{k}$,
 spin index $\lambda=\up,\dwn$  and  isospin index $i$=p, n.
  The Clebsch-Gordan coefficients  in the spin and isospin spaces lead to the projection of the 
  pair operator to appropriate channels.

 The effective Hamiltonian in these pairing channels can be written as 
\beq
\label{eq00}
H&=&\sum_{\bm{p}}\sum_{\lambda=\up,\dwn}\sum_{i={\rm p},{\rm n}}\xi_{\bm{p},i} c_{\bm{p},\lambda,i}^{\dag} c^{ }_{\bm{p},\lambda,i} \nonumber \\
&+& \frac{1}{2} \sum_{\bm{k},\bm{k}',\bm{q}} 
   \left[  \sum_{m=-1}^{+1}  
S^{\dagger}_m (\bm{k},\bm{q})  V_{\rm s}(\bm{k},\bm{k}') S_m (\bm{k}',\bm{q}) 
  +  \sum_{\ell=-1}^{+1}   T^{\dagger}_\ell (\bm{k},\bm{q}) V_{\rm t}(\bm{k},\bm{k}')  T_\ell (\bm{k}',\bm{q}) \right] ,
\eeq
 where $V_{\rm s(t)}$ is a spin-singlet (triplet) interaction
 as functions of the momentums, $\bm{k}$ and $\bm{k}'$. 
 $\xi_{\bm{p},i}=\frac{\bm{p}^2}{2M_{i}}-\mu_{i}$ is the kinetic energy,
measured from the nucleon chemical potentials~$\mu_{i}$.
$M_{i}$ is the nucleon mass.   
The explicit form of  Eq.(\ref{eq00}) is given by
\beq
\label{eq1}
H&=&\sum_{\bm{p}}\sum_{\sigma=\up,\dwn}\sum_{i={\rm n},{\rm p}}\xi_{\bm{p},i}c_{\bm{k},\s,i}^{\dag}c_{\bm{p},\s,i} 
\cr
& {+}&\sum_{\bm{k},\bm{k}',\bm{q}}\sum_{i={\rm n, p}}V_{\rm s}(\bm{k},\bm{k}')
c_{\bm{  k}+\bm{q}/2,\up,i}^{\dag}c_{-\bm{  k}+\bm{q}/2,\dwn,i}^{\dag}
c_{-\bm{k}'+\bm{q}/2,\dwn,i}c_{\bm{k}'+\bm{q}/2,\up,i}
 \cr
&+&
\sum_{\bm{k},\bm{k}',\bm{q}}\sum_{\s=\up,\dwn} V_{\rm t}(\bm{k},\bm{k}')
c_{\bm{  k}+\bm{q}/2,\s,{\rm n}}^{\dag}c_{-\bm{  k}+\bm{q}/2,\s,{\rm p}}^{\dag}
c_{-\bm{k}'+\bm{q}/2,\s,{\rm p}}c_{\bm{k}'+\bm{q}/2,\s,{\rm n}}
\cr
& {+}&\frac{1}{2}\sum_{\bm{k},\bm{k}',\bm{q}} V_{\rm s}(\bm{k},\bm{k}')
\Bigl[c_{\bm{  k}+\bm{q}/2,\up,{\rm n}}^{\dag}c_{-\bm{  k}+\bm{q}/2,\dwn,{\rm p}}^{\dag}
+c_{\bm{k}+\bm{q}/2,\up,{\rm p}}^{\dag}c_{-\bm{k}+\bm{q}/2,\dwn,{\rm n}}^{\dag}
\Bigr]\cr
&&\ \ \ \ \ \ \ \ \ \ \ \ \ \ \ \ \ \ \  \ \ \ \ \ \  \times 
\Bigl[c_{\bm{k}'+\bm{q}/2,\up,{\rm n}}c_{-\bm{k}'+\bm{q}/2,\dwn,{\rm p}}
+c_{\bm{k}'+\bm{q}/2,\up,{\rm p}}c_{-\bm{k}'+\bm{q}/2,\dwn,{\rm n}}
\Bigr]
\cr
& {+}&\frac{1}{2} \sum_{\bm{k},\bm{k}',\bm{q}} V_{\rm t}(\bm{k},\bm{k}')
 \Bigl[ { c_{\bm{  k}+\bm{q}/2,\up,{\rm n}}^{\dag}c_{-\bm{  k}+\bm{q}/2,\dwn,{\rm p}}^{\dag}}
-c_{\bm{k}+\bm{q}/2,\up,{\rm p}}^{\dag}c_{-\bm{k}+\bm{q}/2,\dwn,{\rm n}}^{\dag}
\Bigr]\cr
&&\ \ \ \ \ \ \ \ \ \ \ \ \ \ \ \ \ \ \ \ \ \ \ \ \ \times \Bigl[ {c_{\bm{k}'+\bm{q}/2,\up,{\rm n}}c_{-\bm{k}'+\bm{q}/2,\dwn,{\rm p}}}
-c_{\bm{k}'+\bm{q}/2,\up,{\rm p}}c_{-\bm{k}'+\bm{q}/2,\dwn,{\rm n}}
\Bigr] .
\eeq

\subsection{Effective $S$-wave Interaction}

Throughout  this paper, we neglect the isospin symmetry breaking in the interaction $V_{\rm s(t) }$ and use the 
  averaged nucleon  mass, $M_{\rm p}=M_{\rm n}=M=939$ MeV.
Furthermore, we only retain the $S$-wave part of  $V_{\rm s(t) }$ at low energies  and 
 introduce a multi-rank separable  potential~\cite{Yamaguchi,Mongan,Mongan2,Mathelitsch,Haidenbauer,Haidenbauer3,Grygorov} 
  \beq
\label{eq2}
V^{\rm SEP}_{\alpha}({k},{k}')
=\sum_{N=1}^{N_{\rm max}}\eta_{\alpha,N}\gamma_{\alpha,N}(k)\gamma_{\alpha,N}(k'),
\eeq
where $\gamma_{\alpha,N}(k) >0$ is a form factor with the suffix $\alpha={\rm s}, {\rm t}$  representing the
spin-singlet ($\alpha={\rm s}$) and spin-triplet ($\alpha={\rm t}$) channels, respectively. 
$\eta_{\alpha,N}=\pm 1$ determines the sign of the interaction  (e.g., $\eta_{\alpha,N}=-1$ is attractive).
We note that the partial wave expansion of the potential reads
$V_{\alpha}(\bm{k},\bm{k}')= 4 \pi \sum_{L,M} V_{\alpha} ^{(L,M)} (k,k')  Y_{L M} (\hat{\bm{k}}) Y_{L M} (\hat{\bm{k}'})$ with
 $\alpha= {\rm s}({\rm t})$.  Eq.(\ref{eq2}) is a separable approximation of the $S$-wave contribution, $V_{\alpha} ^{(0,0)} (k,k')$. 
Such a separable potential has been successfully applied to various nuclear systems~\cite{Pieter1,Osman,Alm,Schnell,Sedrakian2,Beyer,Schadow,Bozek,Dewulf,Stein,Jin,Martin}.   

\begin{figure}[t]
\begin{center}
\includegraphics[width=13.5cm]{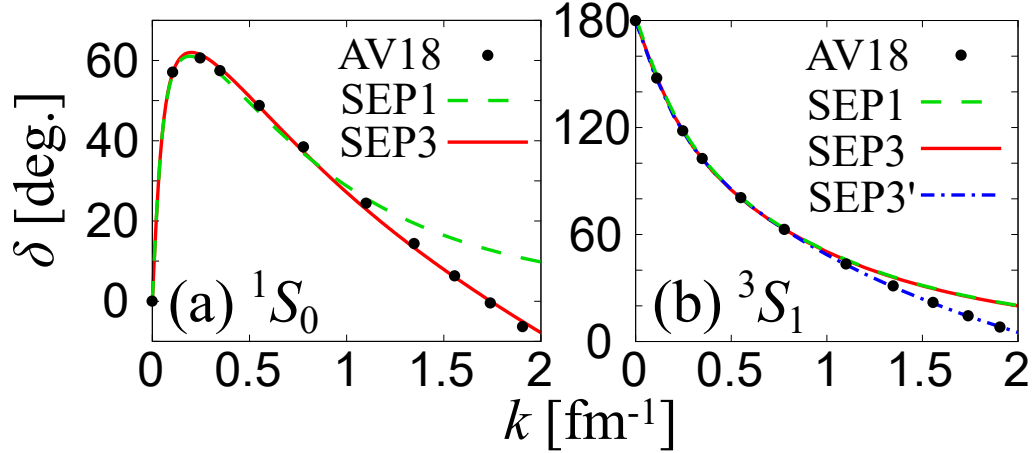}
\end{center}
\caption{Phase shifts of (a) $^1S_0$ neutron-neutron and (b) $^3S_1$ neutron-proton interaction. 
In each figure, black dots show the AV18 phase shift in Ref. \cite{AV18}.
SEP1 and SEP3 represent results of the rank-one and rank-three separable potentials, respectively.
}
\label{fig1}
\end{figure}

The simplest case is the rank-one separable  potential (SEP1), which is given by setting $j_{\rm max}=1$ and  $\eta_{\alpha,1}=-1$ in Eq. (\ref{eq2}).
A typical example of SEP1 is the Yamaguchi potential \cite{Yamaguchi}, 
\beq
\label{eq3}
V_{\alpha}^{\rm  SEP1}({k},{k}')
=\eta_{\alpha,1}\gamma_{\alpha,1}({k})\gamma_{\alpha,1}({k}')
=-\frac{u_{\alpha,1}}{k^2+\Lambda_{\alpha,1}^2}\frac{u_{\alpha,1}}{k'^2+\Lambda_{\alpha,1}^2}.
\eeq
The parameters $u_{\alpha,1}$ and $\Lambda_{\alpha,1}$ are determined such that
the observed values of the scattering length and the effective range in the $^1S_0$ channel ($a_{\rm s}, r_{\rm s}$)=(-18.5 fm, 2.80 fm), 
 and those in the $^3S_1$ channel ($a_{\rm t}, r_{\rm t}$)=(5.42 fm, 1.76 fm)  
can be reproduced:
\beq
\label{eq4}
u_{\alpha,1}=\Lambda_{\alpha,1}^2\sqrt{\frac{8\pi}{M}\frac{1}{\Lambda_{\alpha,1}-2/a_{\alpha}}},
\quad
\Lambda_{\alpha,1}=\frac{3+\sqrt{9-16r_{\alpha}/a_{\alpha}}}{2r_{\alpha}}.
\eeq
We summarize the evaluated values of $u_{\alpha,1}$ and $\Lambda_{\alpha,1}$ in Table I, as well as
 the resulting phase shifts denoted by the dashed lines in  Fig. \ref{fig1}(a,b).
 The filled black circles in the figure represent the empirical phase shifts obtained from the 
  high-precision phenomenological potential, AV18 \cite{AV18}.
    In the  low-momentum region ($k \lesssim 1\ \rm{fm}^{-1}$), a reasonable agreement between SEP1 and AV18 is
    obtained in both $^1S_0$ and $^3S_1$ channels, while substantial deficit of the repulsion 
is seen in the high-momentum region, $k\gesim 1$ fm$^{-1}$ in both channels.
 
  A better agreement with AV18 in the high momentum region is obtained in
 the rank-three separable potential  (SEP3), which is given by setting $j_{\rm max}=3$, 
$(\eta_{\alpha,1},\eta_{\alpha,2},\eta_{\alpha,3})=(-1,1,1)$ and the form factors as,
\beq
\label{eq5}
\gamma_{\alpha,1}=\frac{u_{\alpha,1}}{k^2+\Lambda_{\alpha,1}^2},\quad
\gamma_{\alpha,2}=\frac{u_{\alpha,2}}{k^2+\Lambda_{\alpha,2}^2},\quad
\gamma_{\alpha,3}=\frac{u_{\alpha,3}k^2}{(k^2+\Lambda_{\alpha,3}^2)^2}.
\eeq
In Table I,  we summarize 
the SEP3 parameters determined so as to reproduce the AV18 phase shifts in the range 0 fm$^{-1}$ $\le$ $k$ $\le$ 2 fm$^{-1}$,
 as well as the empirical scattering lengths and effective ranges. 
As shown in  Fig. \ref{fig1}(a), the SEP3 potential (the red line) well reproduces the $^1S_0$  phase shift $\delta$, even beyond  $k\simeq 1.75$ fm$^{-1}$, where $\delta$ turns into negative.  On the other hand, the SEP3 potential overestimates the phase shift $\delta$ in the $^3S_1$ channel (the red line)
in Fig. \ref{fig1}(b) when $k\gesim 1$ fm$^{-1}$.
  
To further improve the agreement, we introduce  a SEP3' potential for the  $^3S_1$ channel with the 
     parameters in TABLE  I.
 Here, the AV18 phase shift is fitted in the range 0 fm$^{-1}$ $\le$ $k$ $\le$ 2 fm$^{-1}$,
  without stringent constraint on the empirical value of $r_{\rm t}$.  Although  the
   effective range and the deuteron binding energy, in SEP3' differ from the empirical values by about 9\% and 4\%, 
   respectively, (see TABLE II),
  one sees in Fig.~\ref{fig1}(b) that SEP3' (blue dash-dotted line) gives good agreement with AV18  
 to $k\simeq 2$ fm$^{-1}$.
 In the following, we employ SEP1, SEP3 and SEP3', to study the superfluid instabilities of asymmetric nuclear matter.

\begin{table}
\caption{Parameters of rank-one (SEP1) and rank-three (SEP3)  separable potentials in $^1S_0$ ($\alpha={\rm s}$) and $^3S_1$ ($\alpha={\rm t}$) channels.}
\begin{ruledtabular}
\begin{tabular}{ccccccc}
 &$u_{\alpha,1}$ [fm$^{-1}$]&$u_{\alpha,2}$ [fm$^{-1}$]&$u_{\alpha,3}$ [fm$^{-1}$]&
 $\Lambda_{\alpha,1}$ [fm$^{-1}$]&$\Lambda_{\alpha,2}$ [fm$^{-1}$]&$\Lambda_{\alpha,3}$ $[{\rm fm}^{-1}]$ \\
\hline
$^1S_0$ ($\alpha={\rm s}$,  SEP1) & 2.6683 & 0 & 0  &1.1392
&  -- & --   \\
$^1S_0$ ($\alpha={\rm s}$,  SEP3) & 4.3097 & 4.5185 & 104.82 &1.3952
& 2.3202  & 3.2578   \\ \hline
$^3S_1$ ($\alpha={\rm t}$, SEP1) & 4.4592 & 0 & 0  &1.4064
& --  & --  \\
$^3S_1$ ($\alpha={\rm t}$,  SEP3)& 4.4619 & 0.1631 & 2.2085 &1.4064
& 2.3455 & 3.0332   \\ 
$^3S_1$ ($\alpha={\rm t}$,  SEP3')& 6.3578 & 1.0956 & 26.814 & 1.7071
& 2.9448 & 2.7045  \\
\end{tabular}
\end{ruledtabular}
\label{table1}
\end{table}
\begin{table}
\caption{Scattering lengths $a_{\alpha}$, effective ranges $r_{\alpha}$, as well as the  binding energy $E_{\rm d}$ of deuteron for $^3S_1$ channel with the parametrization shown in TABLE I.}
\begin{ruledtabular}
\begin{tabular}{cccc}
 &$a_{\alpha}$ [fm] &$r_{\alpha}$ [fm]&$E_{\rm d}$ [MeV] \\
\hline
$^1S_0$ ($\alpha={\rm s}$,  SEP1) & -18.50 & 2.80 &  --   \\
$^1S_0$ ($\alpha={\rm s}$,  SEP3) & -18.50 & 2.80 &  --   \\ \hline
$^3S_1$ ($\alpha={\rm t}$,  SEP1)& \hspace{1.75mm}  5.42 & 1.76 & -2.22   \\
$^3S_1$ ($\alpha={\rm t}$,  SEP3)& \hspace{1.75mm}  5.42 & 1.76 & -2.22   \\ 
$^3S_1$ ($\alpha={\rm t}$,  SEP3')& \hspace{1.75mm}  5.42 & 1.91 & -2.15   \\ 
\end{tabular}
\end{ruledtabular}
\label{table2}
\end{table}


\begin{figure}[t]
\begin{center}
\includegraphics[width=10cm]{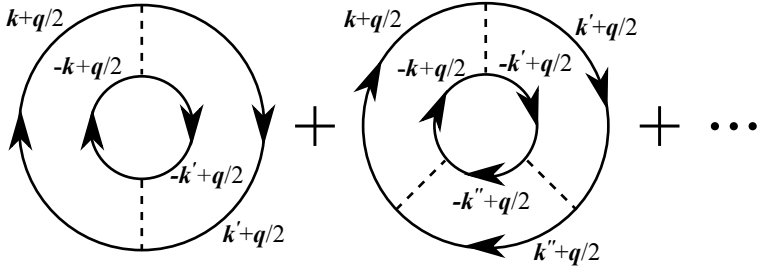}
\end{center}
\caption{
NSR strong-coupling corrections $\delta \Omega_{\rm NSR}$ to the thermodynamic potential $\Omega$ in asymmetric nuclear matter at nonzero temperatures.
The solid and dashed lines denote the nucleon Green's function $G_{i}$ and the bare nucleon-nucleon interaction $V_{\alpha}(\bm{k},\bm{k}^{'})$, respectively. $\bm{k}$, $ \bm{k}^{'}$, and $\bm{k}^{''}$ are relative momenta of nucleons and $\bm{q}$ is the center-of-mass momentum of each pair.}
\label{fig2}
\end{figure}

\subsection{Thermodynamic Potential with Pairing Fluctuations}
We include strong pairing fluctuations originating from $V_{\alpha={\rm s,t}}$ at finite temperatures
within the framework of NSR.
In this scheme, the so-called strong-coupling corrections $\delta \Omega_{\rm NSR}$
to the thermodynamic potential $\Omega$ 
are
 diagrammatically given in Fig. \ref{fig2}.
We note that effects of pairing fluctuations for 
 pure neutron matter at zero temperature was previously discussed in \cite{Pieter1}  by using  a rank-one separable interaction.
Considering the spin-unpolarized nuclear matter,
we introduce the one-particle thermal Green's function in the Hartree approximation,
given by
 \beq
\label{eq11}
G_{\bm{p},i}(i\omega_l)=\frac{1}{i\omega_l-\xi_{\bm{p},i}-\Sigma_{i}^{\rm H}(\bm{p})}.
\eeq
Here, the Hartree self-energy $\Sigma_{i}^{\rm H}(\bm{p})$ involves the contribution from the diagonal force $V_{\rm D}^{\rm SEP} (k,k) $ in the isospin space originating from the
 nn and pp interactions, as well as that from the off-diagonal force  $V_{\rm OD}^{\rm SEP} (k,k) $ originating from the np interactions: 
\beq
\label{eq12}
\Sigma_{i}^{\rm H}(\bm{p}) &=&T\sum_{\bm{p'},\omega_l} \left[  V_{\rm D}^{\rm SEP} (k,k)    G_{\bm{p'},i}(i\omega_l)
+ V_{\rm OD}^{\rm SEP} (k,k) G_{\bm{p}', \bar{i}}(i\omega_l)\right], \\
  &  &V_{\rm D}^{\rm SEP} (k,k)  =    V_{\rm s}^{\rm SEP} (k,k),  \\
  &  &  V_{\rm OD}^{\rm SEP} (k,k) =  V_{\rm t}^{\rm SEP} (k,k) 
    + \frac{1}{2} \left[   V_{\rm s}^{\rm SEP} (k,k)  + V_{\rm t}^{\rm SEP} (k,k)  \right] ,
\eeq
where $\bar{i}$=p(n) for ${i}$=n(p),  $k=|\bm{p}-\bm{p'}|/2$, and  $\omega_l=(2l+1)\pi T$ is the fermion Matsubara frequency.

Introducing the Fermi momentum distribution for given momentum $\bm{p}$ in the Hartree approximation,
\beq
\label{eq13-1}
\rho_{\bm{p},i}^{\rm H}  &=&  T\sum_{\omega_l}G_{\bm{p},i}(i\omega_l), 
 \eeq
one can write the thermodynamic potential $\Omega$ in the NSR theory as,
\beq
\label{eq6}
\Omega \ &=& \Omega_{\rm H} + \delta \Omega_{\rm NSR}, \cr
 \Omega_{\rm H}  & = &  2 T   \sum_{\bm{p},i}{\rm ln}\left[1+e^{-\xi_{\bm{p},i}^{\rm H}/T}\right] 
 - \sum_{\bm{p},\bm{p}',i}  \left[   V_{\rm D}^{\rm SEP} (k,k) \rho_{\bm{p},i}^{\rm H} \rho_{\bm{p'},i}^{\rm H}
         + V_{\rm OD}^{\rm SEP} (k,k) \rho_{\bm{p},i}^{\rm H} \rho_{\bm{p'},\bar{i}}^{\rm H} \right],  \cr
 \delta \Omega_{\rm NSR} &=&
 T \sum_{\bm{q}, \nu_l}  \sum_{\alpha} \sum_{m=0,\pm 1} {\rm Tr}\Bigl[{\rm ln}\left[1+\hat{\eta}_{\alpha }\hat{\Pi}^{(m)}_{\alpha}(\bm{q},i\nu_l)\right]-\hat{\eta}_{\alpha}\hat{\Pi}^{(m)}_{\alpha}(\bm{q},i\nu_l)\Bigr] .
\eeq
Here, $\xi_{\bm{p},i}^{\rm H}=\frac{\bm{p}^2}{2M}-\mu_{i}+\Sigma_{i}^{\rm H}(\bm{p})$ is the kinetic energy involving the Hartree self-energy $\Sigma_{i}^{\rm H}(\bm{p})$, measured from the chemical potential $\mu_{i}$, and $\nu_l=2\pi lT$ is  the boson Matsubara frequency.
$\delta\Omega_{\rm NSR}$ in Eq. (\ref{eq6}) is the strong-coupling correction 
to $\Omega$ associated with pairing fluctuations in the $^1S_0$ and $^3S_1$ channels
, and 
 $\hat{\eta}_{\alpha}={\rm diag}(\eta_{\alpha,1},\eta_{\alpha,2},...,\eta_{\alpha,N_{\rm max}})$
Note that  ${\rm Tr}$ is to take over the rank indices, $N$.
The $N_{\rm max} \times N_{\rm max} $ matrix pair-correlation function  $\hat{\Pi}_{\alpha}^{(m)}(\bm{q},i\nu_l)=\{[\Pi_{\alpha}^{(m)}(\bm{q},i\nu_l)]_{N,N'}\}$ consists of
 \beq
\label{eq7}
\left[\hat{\Pi}_{\rm s}^{(+1)}(\bm{q},i\nu_l)\right]_{N,N'}&=&T\sum_{\bm{k},\omega_{l'}}\gamma_{{\rm s},N}(k)\gamma_{{\rm s},N'}(k)G_{\bm{k}+\bm{q}/2,{\rm p}}(i\omega_{l'}+i\nu_l)G_{-\bm{k}+\bm{q}/2,{\rm p}}(-i\omega_{l'}),
\eeq
\beq
\label{eq8}
\left[\hat{\Pi}_{\rm s}^{(0)}(\bm{q},i\nu_l)\right]_{N,N'}\ &=&T\sum_{\bm{k},\omega_{l'}}\gamma_{{\rm s},N}(k)\gamma_{{\rm s},N'}(k)G_{\bm{k}+\bm{q}/2,{\rm n}}(i\omega_{l'}+i\nu_l)G_{-\bm{k}+\bm{q}/2,{\rm p}}(-i\omega_{l'}),
\eeq
\beq
\label{eq9}
\left[\hat{\Pi}_{\rm s}^{(-1)}(\bm{q},i\nu_l)\right]_{N,N'}&=&T\sum_{\bm{k},\omega_{l'}}\gamma_{{\rm s},N}(k)\gamma_{{\rm s},N'}(k)G_{\bm{k}+\bm{q}/2,{\rm n}}(i\omega_{l'}+i\nu_l)G_{-\bm{k}+\bm{q}/2,{\rm n}}(-i\omega_{l'}),
\eeq
\beq
\label{eq10}
\left[\hat{\Pi}_{\rm t}^{(0,\pm 1)}(\bm{q},i\nu_l)\right]_{N,N'}&=&T\sum_{\bm{k},\omega_{l'}}\gamma_{{\rm t},N}(k)\gamma_{{\rm t},N'}(k)G_{\bm{k}+\bm{q}/2,{\rm n}}(i\omega_{l'}+i\nu_l)G_{-\bm{k}+\bm{q}/2,{\rm p}}(-i\omega_{l'}),
\eeq
where $N, N'=1,2,...,N_{\rm max}$.

Since we are considering the spin-unpolarized case, Eqs. (\ref{eq7})-(\ref{eq10}) are spin-independent.
We briefly note that the first order correction ${\rm Tr} [\hat{\eta}_{\alpha}\hat{\Pi}^{(m)}_{\alpha}(\bm{q},i\nu_l)] $
 is already involved in 
the Hartree self-energy $\Sigma_{i}^{\rm H}(\bm{p})$ \cite{Pieter1}, so that 
we have removed it in Eq.(\ref{eq6}) to avoid double counting.
 
\subsection{Critical Temperature}

The critical temperatures of the $^1S_0$ neutron superfluidity ($T_{\rm c}^{\rm nn}$), $^1S_0$ proton superconductivity ($T_{\rm c}^{\rm pp}$) and $^3S_1$ deuteron condensation
($T_{\rm c}^{\rm d}$), as functions of 
baryon density are obtained  
 from the Thouless criterion \cite{Thouless}.
Here, we introduce the Thouless determinant  $D_{\alpha}^{(m)} (T)$ defined by 
\beq
\label{eq13}
D_{\rm s}^{(-1)} (T) & \equiv &  {\rm det}\bigl[1+\hat{\eta}_{\rm s}\hat{\Pi}_{\rm s}^{(-1)}(\bm{q}=0,i\nu_l=0)\bigr] =0 \ \ {\rm at} \ T=T_{\rm c}^{\rm nn},\\
\label{eq14}
D_{\rm s}^{(+1)} (T) & \equiv &  {\rm det}\bigl[1+\hat{\eta}_{\rm s}\hat{\Pi}_{\rm s}^{(+1)}(\bm{q}=0,i\nu_l=0)\bigr]=0 \ \ {\rm at} \ T=T_{\rm c}^{\rm pp}, \\
\label{eq141}
D_{\rm t}^{(0,\pm 1)} (T)  & \equiv &  {\rm det}\bigl[1+\hat{\eta}_{\rm t}\hat{\Pi}_{\rm t}^{(0,\pm 1)}(\bm{q}=0,i\nu_l=0)\bigr]=0  \ \ {\rm at} \ T=T_{\rm c}^{\rm d}.
\eeq
We briefly note that Eqs. (\ref{eq13})-(\ref{eq141}) originate from a ``block diagonalized'' matrix pair-correlation function with respect to $m=\pm1,0$, so that the Thouless criterion is decomposed into the three equations (\ref{eq13})-(\ref{eq141}).
We actually solve them, together with the particle number equation for the nucleon density, 
\beq 
 \label{eq15}
\rho_{i}=-\frac{\partial \Omega}{\partial \mu_{i}}.
\eeq

In this paper, we approximate  $\Sigma_{i}^{\rm H}(\bm{p})$ to the value at the 
 Fermi surface (for theoretical backfround, see Appendix \ref{appA}).  Then,
we have
\beq
\label{eq16}
\Sigma_{i}^{\rm H}(\bm{p})\simeq \Sigma_{i}^{\rm H}(\bm{p}=\bm{k}_{{\rm F},i})\equiv \overline{\Sigma}_{i}^{\rm H},
\eeq
where $k_{{\rm F},i}$ is the nucleon Fermi momentum.
Introducing the effective chemical potential 
\beq
\label{eq161}
 \mu_i^{\rm H} \equiv \mu_i - \overline{\Sigma}_{i}^{\rm H},
\eeq
one can write the particle number equation in the form,
\beq
\label{eq17}
 \rho_{i}=\rho_{i}^{\rm H}+ \sum_{i'}\delta\rho_{i'}^{\rm NSR} L_{i'i}, 
\eeq
where the Hartree density $\rho_{i}^{\rm H}$ and the NSR correction $\delta\rho_{i}^{\rm NSR}$ are, respectively, given by
\beq
\label{eq18}
\rho_{i}^{\rm H}= 2 \sum_{\bm p} \rho_{\bm{p},i}^{\rm H},  \ \
 \delta\rho_{i}^{\rm NSR} = - \frac{\partial ({\delta \Omega_{\rm NSR}})}{\partial \mu_{i}^{\rm H}}.
\eeq
The NSR correction $\delta\rho_{i}^{\rm NSR}$ to the number equation involves the diagonal and off-diagonal component of the matrix, 
\beq
L_{ij}=\delta_{ij}-\frac{\partial\overline{\Sigma}_{i}^{\rm H}}{\partial \mu_{j}}.
\eeq
This correction naturally arises from $\delta\Omega_{\rm NSR}$, whereas it was ignored in the previous work~\cite{Ramanan,Jin,Alm,Stein}.
We note that $L_{ij}$ is related to the compressibility matrix $K_{ij}^{\rm H}$ in the mean-field approximation as
\begin{eqnarray}
K_{ij}^{\rm H} \equiv \frac{\partial\rho_{i}^{\rm H}}{\partial\mu_{j}}=-T\sum_{\bm{p},\omega_l}\left[G_{\bm{p},i}(i\omega_n)\right]^2L_{ij},
\end{eqnarray}
which indicates that $L_{ij}$ corresponds to the vertex correction to the density correlation function.
The explicit form of $L_{ij}$ is given by
\beq
\label{eq19}
\left(\begin{array}{cc}
L_{\rm nn} & L_{\rm np} \\
L_{\rm pn} & L_{\rm pp} 
\end{array}
\right)
=\frac{1}{(1+\kappa_{\rm n})(1+\kappa_{\rm p})-\chi_{\rm n}\chi_{\rm p}}
\left(\begin{array}{cc}
1+\kappa_{\rm p} & -\chi_{\rm p} \\
-\chi_{\rm n} & 1+\kappa_{\rm n} 
\end{array}
\right),
\eeq
where
\beq
\label{eq20}
\kappa_{i}&=&-T\sum_{\bm{p},\omega_l}V_{\rm D}^{\rm SEP}(\bar{k},\bar{k})\left[G_{\bm{p},i}(i\omega_n)\right]^2, \\
\label{eq21}
\chi_{i} &=& -T\sum_{\bm{p},\omega_l}V_{\rm OD}^{\rm SEP}(\bar{k},\bar{k})\left[G_{\bm{p},i}(i\omega_n)\right]^2,
\eeq
with $\bar{k}=|\bm{k}_{{\rm F},i}-\bm{p}|/2$.

The asymmetric nuclear matter can conveniently be characterized by the 
total baryon density $\rho$ and the proton fraction $Y_{\rm p}$, respectively given by
\beq
\rho=\rho_{\rm n}+\rho_{\rm p}, \ \ \ \ Y_{\rm p}=\frac{\rho_{\rm p}}{\rho_{\rm n}+\rho_{\rm p}}.
\eeq
 Below, we treat  $\rho$ and $Y_{\rm p}$ as independent parameters, to study their effects on the 
 critical temperatures, $T_{\rm c}^{\rm nn}$, $T_{\rm c}^{\rm d}$, and $T_{\rm c}^{\rm pp}$. 
We briefly note that, in real neutron star matter,  the charge neutrality as well as the chemical equilibrium conditions among protons, neutrons, electrons and muons
 provide a constraint between $\rho$ and $Y_{\rm p}$~\cite{APR}.

 \section{Results}
\label{sec3}
\begin{figure}[t]
\begin{center}
\includegraphics[width=13.5cm]{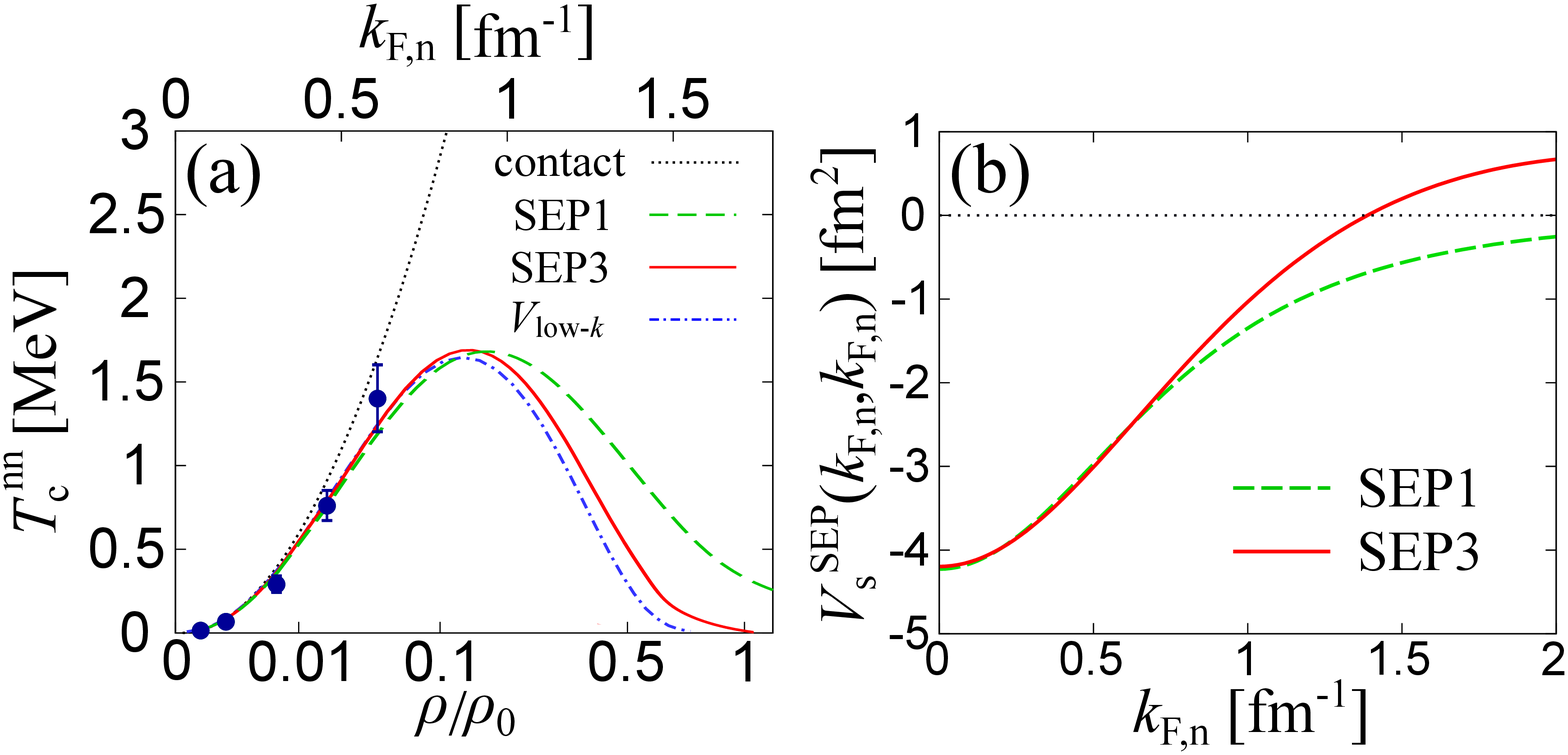}
\end{center}
\caption{(a) Calculated $^{1}S_0$ neutron superfluid phase transition temperature $T_{\rm c}^{\rm nn}$ as a function of a nucleon density $\rho=\rho_{\rm n}$ in pure neutron matter.
$k_{\rm F,n}=(3\pi^2\rho_{\rm n})^{1\over 3}$ is the neutron Fermi momentum.
The dotted, dashed, and solid lines denote the NSR results of the contact-type (``contact"), rank-one separable (``SEP1"), and rank-three separable (``SEP3") interactions, respectively.
 ``$V_{{\rm low}-k}$" (dot-dashed line) corresponds to the previous NSR work of the renormalization-group based low-momentum interaction~\cite{Ramanan}.
The filled circles represent the result of the lattice Monte-Carlo simulation for the pionless effective 
theory~\cite{Abe}.
(b) The strength of the nn interaction on the Fermi surface, as a function of the neutron  Fermi momentum.}
\label{fig3and4}
\end{figure}
We start from the superfluid phase transition temperature
$T_{\rm c}^{\rm nn}$  in pure neutron matter  ($Y_{\rm p}=0$) 
 which has been  studied in  different levels of theoretical sophistication before.
Figure~\ref{fig3and4} (a) shows theoretical estimates of $T_{\rm c}^{\rm nn}$~\cite{note}.
The NSR result of the rank-three separable potential (``SEP3") shows good agreement with the previous work of NSR with an effective low-momentum interaction $V_{{\rm low}-k}$ based on the renormalization group~\cite{Ramanan}, as well as
the result of the lattice Monte-Carlo simulations for the pionless effective field theory~\cite{Abe} shown by the filled circle (where the interaction is chosen so as to reproduce the nn scattering length and the nn effective range).
\par 
To see effects of the effective range and the short-range repulsion in the $^1S_0$ nn channel, we also plot in Fig.~\ref{fig3and4} (a) the calculated $T_{\rm c}^{\rm nn}$ of NSR with the contact-type interaction $V_{\rm s}(k,k')=u_{\rm s,1}^2$ (``contact''), where $u_{\rm s,1}$ is chosen so as to reproduce $a_{\rm s}$, 
and the rank-one separable interaction (``SEP1'').
In the low-density regime ($\rho/\rho_0 < 0.01$) including the neutron drip density $\rho_{\rm drip}/\rho_0\simeq 1.5\times10^{-3}$ \cite{Dean},  all four theoretical calculations agree well with each other and with the
 Monte Carlo data, indicating that the critical temperature is determined  only by  the scattering length.
 The non-zero effective range ($r_{\rm s}=2.8$ fm) suppresses $T_{\rm c}^{\rm nn}$ when $\rho/\rho_0\gesim 0.1$ [see Fig. \ref{fig3and4} (a)]. 
 It can also be understood as effects of the momentum cut-off $\Lambda_{\rm s,1}$ associated with the effective range \cite{Pieter1,Andrenacci}.
In such a region, the Thouless criterion is approximately given by
\begin{eqnarray}
\label{eq33}
1\simeq V_{\rm s}^{\rm SEP}(k_{\rm F,n},k_{\rm F,n})\sum_{\bm{k}}\frac{1}{2\xi_{\bm{k},{\rm n}}}\tanh\left(\frac{\xi_{\bm{k},{\rm n}}}{2T_{\rm c}^{\rm nn}}\right).
\end{eqnarray}
From Eq.~(\ref{eq33}), one can find that the nn interaction strength on the Fermi surface $V_{\rm s}^{\rm SEP}(k_{\rm F,n},k_{\rm F,n})$ is of importance to evaluate $T_{\rm c}^{\rm nn}$.
Figure~\ref{fig3and4} (b) shows $V_{\rm s}^{\rm SEP}(k_{\rm F,n},k_{\rm F,n})$ of SEP1 and SEP3.
Since $V_{\rm s}^{\rm SEP}(k,k')$ of SEP1 and SEP3 are given by Eqs.~(\ref{eq3}) and (\ref{eq5}), respectively, 
$V_{\rm s}^{\rm SEP}(k_{\rm F,n},k_{\rm F,n})$ decreases with increasing $k_{\rm F,n}$.
The decrease of $V_{\rm s}^{\rm SEP}(k_{\rm F,n},k_{\rm F,n})$ is associated with $\Lambda_{\rm s,1}\simeq 3/2r_{\rm s}$.
We briefly note that such a decrease does not occur in the case of the contact-type interaction which is momentum-independent. 
Moreover, the short-range repulsion of the nn interaction 
 takes over for $\rho/\rho_0 > 0.54$ (near the crust-core transition density $\rho/\rho_0\sim 0.5$~\cite{Page3}) to further suppress $T_{\rm c}^{\rm nn}$ as $V_{{\rm low-}k}$ and SEP3 shown in Fig.\ref{fig3and4} (a).
  Indeed, the comparison of SEP1 and SEP3 interactions  on the Fermi surface $V_{\rm s}^{\rm SEP}(k_{\rm F,n},k_{\rm F,n})$ 
  shown in Fig.~\ref{fig3and4} (b) indicates that the typical strength of the nn interaction decreases with increasing neutron density, and
turns into repulsive for $k_{\rm F,n} > 1.39~{\rm fm}^{-1}$.
Good agreement of our SEP3 result with the previous  $V_{{\rm low-}k}$ result over the wide range of baryon density
indicates the importance of the detailed interaction structure, as well as associated pairing fluctuations to obtain $T_{\rm c}^{\rm nn}$.  
\par
\begin{figure}[t]
\begin{center}
\includegraphics[width=7cm]{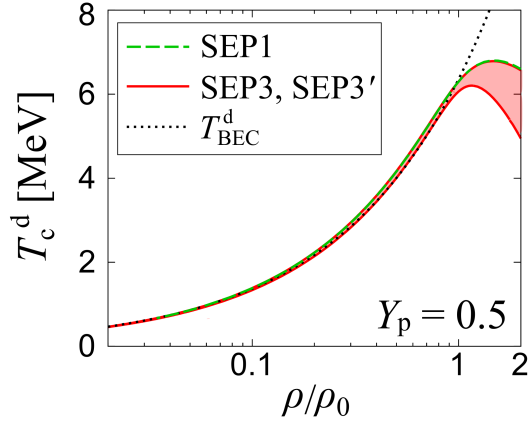}
\end{center}
\caption{The deuteron condensation temperature $T_{\rm c}^{\rm d}$ in the $^3S_1$ channel in symmetric nuclear matter ($Y_{\rm p}=0.5$).
The upper and lower bounds of the solid band correspond to the results using the parameter sets shown in Tables I and II, that is, SEP3 and SEP3', respectively. 
$T_{\rm BEC}^{\rm d}$ shows the Bose-Einstein condensation temperature of deuteron gases where the deuteron is assumed as a noninteracting boson.
}
\label{fig5}
\end{figure}

We proceed to the case of the symmetric nuclear matter ($Y_{\rm p}=0.5$).
  In this case, examining the Thouless criterion for the nn, pp and np pairing channels, 
we find that the highest critical temperature is always obtained in the deuteron np channel to $\rho/\rho_0\leq2$.
Figure~\ref{fig5} shows the critical temperature of the deuteron condensation, $T_{\rm c}^{\rm d}$ obtained by
  SEP3 and SEP3' for np interaction with SEP3 for nn and pp interactions.
The upper (lower) bound of the red solid band corresponds to SEP3 (SEP3').
The green dashed line represents the result of SEP1.
  For comparison, we also plot in Fig.~\ref{fig5} the Bose-Einstein condensation temperature of an assumed noninteracting deuteron gas, given by~\cite{Stein,Tajima3,Jin}
\beq
\label{eqdBEC}
T_{\rm BEC}^{\rm d}=\frac{\pi}{m}\left[\frac{\rho_{\rm n}}{3\zeta(3/2)}\frac{Y_{\rm p}}{1-Y_{\rm p}}\right]^{2\over 3}.
\eeq
The obtained $T_{\rm c}^{\rm d}$ with all separable interaction potentials approaches $T_{\rm BEC}^{\rm d}$ in the low-density region.
While our result for the symmetric case ($Y_{\rm p}=0.5$) is qualitatively consistent with the previous work using different separable interactions within the NSR framework~\cite{Stein,Jin},
 $T_{\rm c}^{\rm d}$ has a peak structure at $\rho_{\rm peak}/\rho_0 > 1 $,
which is in contrast to the previous work giving
  $\rho_{\rm peak}/\rho_0 = 0.3 - 0.8 $~\cite{Stein,Jin}.
 In addition, we do not find a strange back bending behavior of  $T_{\rm c}^{\rm d}$ 
  seen in \cite{Stein,Jin}, irrespective of the use of SEP1, SEP3 and SEP3'.
  We have not fully understood those differences. 
  However, the treatment of the single-particle energy might be a  possible origin.

\par
\begin{figure}[t]
\begin{center}
\includegraphics[width=7cm]{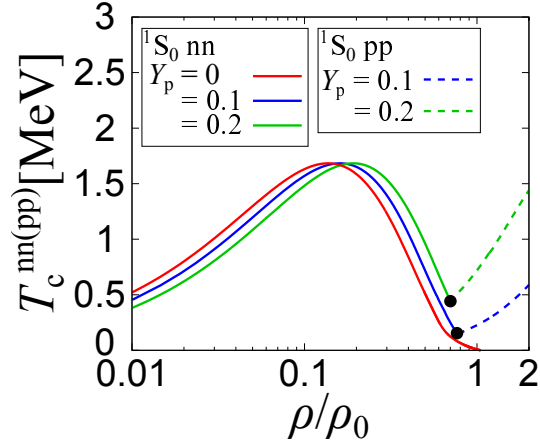}
\end{center}
\caption{
Calculated critical temperatures $T_{\rm c}^{\rm nn}$ (solid) and $T_{\rm c}^{\rm pp}$ (dashed) for $^{1}S_0$ neutron superfluid and proton superconductivity.
The circles represent the nucleon densities where both superfluid instabilities simultaneously occur.
}
\label{fig6}
\end{figure}
We now consider asymmetric nuclear matter within the same theoretical framework.
We restrict ourselves to the case with the low proton fraction, $Y_{\rm p}=0.1\sim 0.2$, (which is, however, still valid to the study of the neutron star cooling 
\cite{Lattimer,APR,Alford}).
 In this range of $Y_{\rm p}$, the absolute value of the relative momentum $k=|\bm{k}|$ between p and n is smaller than 1.29~fm$^{-1}$,
 so that
  we use SEP3 (which gives better agreement with the empirical phase shift at low energies.
  The Thouless criterion for the nn, pp and np channels gives the highest critical temperature in the  nn channel at low densities,
   while the pp pairing takes over above the nuclear matter density. 
 Note here that, in the low-density limit, 
 $T_{\rm BEC}^{\rm d}$ becomes dominant even in asymmetric nuclear matter $0<Y_p<0.5$ (see Appendix \ref{appB}).  
 The deuteron pairing is remarkably suppressed due to imbalanced Fermi surfaces. 
  Figure~\ref{fig6} shows $T_{\rm c}^{\rm nn}$ and $T_{\rm c}^{\rm pp}$ in the case of SEP3~\cite{note}. 

In Fig.~\ref{fig6}, with increasing the proton fraction $Y_{\rm p}$, the peak of $T_{\rm c}^{\rm nn}$ is found to gradually move to higher density.
 This is simply because the neutron density decreases as $\rho_{\rm n} = (1-Y_{\rm p}) \rho$, so that the  whole curve of $T_{\rm c}^{\rm nn}$
  shifts to the right.
  The black circle  in Fig. \ref{fig6} indicates
the density at which $T_{\rm c}^{\rm pp}$ exceeds $T_{\rm c}^{\rm nn}$ when $Y_{\rm p} > 0$.
   Beyond this,  the pp interaction becomes more attractive, due to relatively small proton Fermi momentum $k_{\rm F,p}=(3\pi^2\rho_{\rm p})^{1\over 3}=(3\pi^2\rho Y_{\rm p})^{1\over 3}$, while the nn interaction  is strongly suppressed by the short-range repulsion
 due to large neutron Fermi momentum $k_{\rm F,n}=(3\pi^2\rho_{\rm n})^{1\over 3}=\left[3\pi^2\rho\left(1-Y_{\rm p}\right)\right]^{1\over 3}$.
 At higher density,
   $T_{\rm c}^{\rm pp}$ would also be suppressed, but it is beyond the applicability of the   present formalism (see Appendix \ref{appB}).

\begin{figure}[t]
\begin{center}
\includegraphics[width=13.5cm]{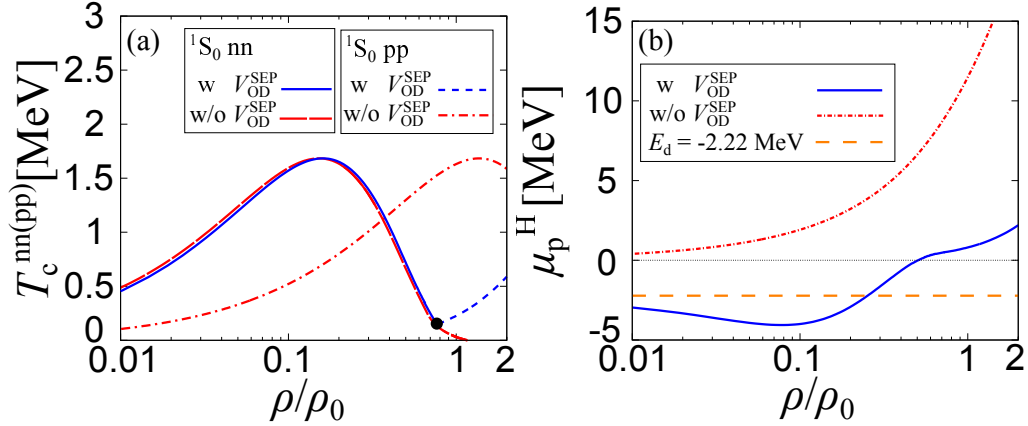}
\end{center}
\caption{(a) The critical temperatures $T_{\rm c}^{\rm nn (pp)}$, and (b) 
the effective proton chemical potential $\mu_{\rm p}^{\rm H}=\mu_{\rm p}-\overline{\Sigma}_{{\rm p}}^{\rm H}$,
 at $Y_{\rm p}=0.1$ with and without the off-diagonal np interaction $V_{\rm OD}^{\rm SEP}$.
The horizontal dashed line in panel (b) represents the deuteron binding energy $E_{\rm d}=-2.22$~MeV.
}
\label{fig7}
\end{figure}

To see effects of  strong np interactions,
we plot the critical temperatures $T_{\rm c}^{\rm nn}$, as well as, $T_{\rm c}^{\rm pp}$ in Fig. \ref{fig7} (a). 
We also show the effective proton chemical potential $\mu_{\rm p}^{\rm H}$ 
which is defined in Eq.(\ref{eq161})
(at $T=T_{\rm c}^{\rm nn}$, below $0.77 \rho_0$ and
  at $T=T_{\rm c}^{\rm pp}$ above $0.77 \rho_0$), 
with and without  the  np interaction, $V_{\rm OD}^{\rm SEP}$
in Figs. \ref{fig7} (b).
We find that while $T_{\rm c}^{\rm nn}$ is insensitive to the strength of the np interaction,  
$T_{\rm c}^{\rm pp}$ is substantially affected. The latter can be understood  
by the  behavior of $\mu_{\rm p}^{\rm H}$.
When $V_{\rm OD}^{\rm SEP}=0$, $\mu_{\rm p}^{\rm H}$ is always positive as shown in Figs. \ref{fig7} (b),
 indicating that the proton Fermi surface is formed, irrespective of the value of baryon density $\rho$, 
 naturally leading to the proton superconductivity.
 On the other hand,  when $V_{\rm OD}^{\rm SEP}\neq 0$,
the strong np interaction in the deuteron channel reduces $\mu_{\rm p}^{\rm H}$ in the low-density region, 
to eventually approach the deuteron binding energy $E_{\rm d}=-2.22$ MeV in the low-density limit.
As a result, the pp pairing does not take place.
In the low density limit with $0 < Y_{\rm p} < 0.5$, one finds
 $\mu_{\rm n} \sim \mu_{\rm n}^{\rm H} \rightarrow 0$ and  $ \mu_{\rm p} \sim
  \mu_{\rm p}^{\rm H}  \rightarrow  E_{\rm d}$~\cite{Tajima3} as in the case of an asymmetric two-component Fermi atomic gas \cite{Liu}.
\begin{figure}[t]
\begin{center}
\includegraphics[width=7cm]{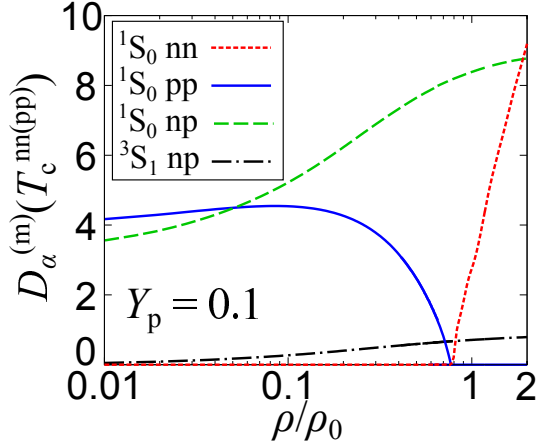}
\end{center}
\caption{Thouless determinants, $D_{\alpha}^{\rm (m)}$ in all four channels as functions of the baryon density $\rho$ with $Y_{\rm p}=0.1$ at $T=T_{\rm c}^{\rm nn}$ below $\rho=0.77\rho_0$ and at $T=T_{\rm c}^{\rm pp}$ above $\rho=0.77\rho_0$.
The dotted, solid, dashed, and dot-dashed lines represent $D_{\alpha}^{\rm (m)}$ of
the $^1S_0$ nn, $^1S_0$ pp, $^1S_0$ np, and $^3S_1$ np channels, respectively. }
\label{fig8}
\end{figure}
\par
Figure~\ref{fig8} shows the Thouless determinants $D_{\alpha}^{(m)}(T)$ in Eqs. (\ref{eq13})-(\ref{eq141}) for $Y_{\rm p}=0.1$ at $T=T_{\rm c}^{\rm nn}$ below $0.77 \rho_0$, and
  at $T=T_{\rm c}^{\rm pp}$ above $0.77 \rho_0$.
  When $D_{\alpha}^{(m)}(T)$ becomes smaller  to vanish, 
  pairing fluctuations become stronger and eventually diverge at the second-order superfluid/superconducting phase transition.
  Such diverging fluctuations can be seen in the $^1S_0$ nn channel for $\rho < 0.77 \rho_0$, as well as 
   in the $^1S_0$ pp channel for $\rho >  0.77 \rho_0$.
On the other hand, pairing fluctuations in the  $^1S_0$ np channel 
    are weak, compared to the other channels. 
   The Thouless determinant in the  $^3S_1$ np channel
   is  close to zero over the entire density, but the deuteron condensation does not occur when $Y_{\rm p}=0.1$, 
    because of the large difference of the chemical potentials between neutrons and protons.  Nevertheless, 
     strong pairing fluctuations in the deuteron channel  play a crucial role for  $T_{\rm c}^{\rm pp}$, as seen 
      in Fig.\ref{fig7}.
\par

\begin{figure}[t]
\begin{center}
\includegraphics[width=7cm]{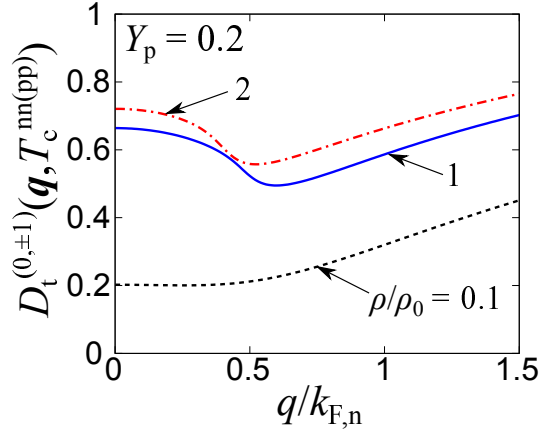}
\end{center}
\caption{Thouless determinant in the deuteron channel as a function of the center-of-mass momentum $q$ at $Y_{\rm p}=0.2$ for
 different values of the baryon density.}
\label{fig9}
\end{figure}

Before ending this section, we discuss the possibility of a Fulde-Ferrell-Larkin-Ovchinnikov (FFLO) state~\cite{FF,LO,Sheehy,Radzihovsky} 
in the deuteron channel for $0 < Y_{\rm p} < 0.2$ (which is relevant for neutron stars).
 The FFLO state may occur, when two kinds of fermions attractively interact with each other in the presence of population imbalance. In such a case, the Cooper pairs with a non-zero center-of-mass momentum are formed.
 In the present case,  the Thouless determinant at a non-zero momentum~\cite{OhashiFF,FrankZwerger}, $D_{\rm t}^{(0,\pm 1)} (\bm{q},T) =   {\rm det}\bigl[1+\hat{\eta}_{\rm t}\hat{\Pi}_{\rm t}^{(0,\pm 1)}(\bm{q},i\nu_l=0)\bigr]$ is an appropriate  measure.
Figure \ref{fig9} shows the center-of-mass momentum ($q=|\bm{q}|$) dependence of $D_{\rm t}^{(0,\pm 1)} (\bm{q},T)$
 at $T=T_{\rm c}^{\rm nn(pp)}$ in asymmetric nuclear matter with $Y_{\rm p}=0.2$.
 We find that $D_{\rm t}^{(0,\pm 1)} (\bm{q},T)$ takes a minimum at a non-zero momentum $q^*$  in the high-density region ($\rho>\rho_0$).
Indeed, $q^*$  at $\rho=\rho_0$  in Fig.\ref{fig9} is close to the typical momentum of the FFLO pairing, 
$k_{\rm F,n}^{\rm eff}-k_{\rm F,p}^{\rm eff}=(2m\mu_{\rm n}^{\rm H})^{1\over 2}-(2m\mu_{\rm p}^{\rm H})^{1\over 2}\simeq 0.7k_{\rm F,n}$.
Although $D_{\rm t}^{(0,\pm 1)} (\bm{q}^*,T)$ is still far away from zero, it may be interpreted as a precursor of the FFLO state at larger $Y_{\rm p}$.

\section{Concluding remarks}
\label{sec4}

In this paper, we have extended the  Nozi\`{e}res-Schmitt-Rink approach to four-component fermion system, to examine
  the  superfluid phase transition at finite temperatures  in asymmetric nuclear  matter at nuclear and subnuclear densities.
 Including pairing fluctuations in  the $S$-wave neutron-neutron, proton-proton, and neutron-proton channels,
 we evaluated the critical temperature of $^1S_0$ neutron superfluidity $T_{\rm c}^{\rm nn}$ 
 and proton superconductivity $T_{\rm c}^{\rm pp}$. 
We  clarified effects of strong neutron-proton pairing fluctuations in the deuteron channel.
 While resultant $T_{\rm c}^{\rm nn}$  in pure neutron matter agrees well with the previous Monte Calro data 
  in the low baryon-density region, it  is remarkably 
   suppressed around the nuclear saturation density  $\rho_0$, due to the short-range nn repulsion.
    We found that $T_{\rm c}^{\rm pp}$ at low-density is substantially suppressed
    by the neutron-proton pairing fluctuations.  
   
There are several future directions to be explored on the basis of the framework developed in this paper.
 \begin{enumerate} 
 \item We have focused on the superfluid/superconducting instability in the normal phase  throughout the paper. However, 
 the present model together with the framework of Ref.~\cite{Pieter1} can be combined 
  to study the superfluid phase below the critical temperature, such as equation of state, as well as magnitude of the pairing gap.
\item To improve the accuracy of  $T_{\rm c}^{\rm nn,pp,d}$, we need to 
  include the coupled $^3S_1$-$^3D_1$ channel potential 
  beyond the present $^3S_1$ channel potential.  Such a channel-coupling introduces 
  extra in-medium effect associated with the Pauli blocking by the intermediate $^3D_1$ state.
 \item There are  correlations which are ignored in the present paper, such as  Gorkov and Melik-Barkhudarov (GMB) screening \cite{GMB,Yu,Pisani},  as well as  
the  competition between the screening and anti-screening corrections~\cite{Cao,Ramanan2}.
 \item The  nn pairing in the $^3P_2$ channel~\cite{Tamagaki,Hoffberg,Takatsuka2,Page3} would cause a
  dominant superfluid component in the liquid core of neutron stars.  Introducing a separable interaction in the 
   $P$-wave channel and applying the present framework would be a first step toward the analysis of such unconventional superfluids.
 \end{enumerate}
 

\par
\acknowledgements
We  thank  
G. Baym, S. Furusawa, S. Han, K. Iida, D. Inotani, T. Kunihiro, H. Liang, P. Naidon, A. Ohnishi, P. Pieri, G. C. Strinati, H. Togashi, and N. Yamamoto for useful discussions. 
H. T. was supported by a Grant-in-Aid for JSPS fellows (No.17J03975).
T. H. was supported by RIKEN iTHEMS Program.
Y. O. was supported by KiPAS project in Keio University. 
This work was supported by Grant-in-aid for Scientific Research from MEXT and JSPS in Japan (No.JP16K17773, No.JP24105006, No.JP23684033, No.JP15H00840, No.JP15K00178, No.JP16K05503, No.JP18H03712, No.JP18H05236, No.JP18H05406, No.JP18K11345, No.JP19K03689). 

\appendix
\section{The Hartree shift}
\label{appA}
\begin{figure}[t]
\begin{center}
\includegraphics[width=7cm]{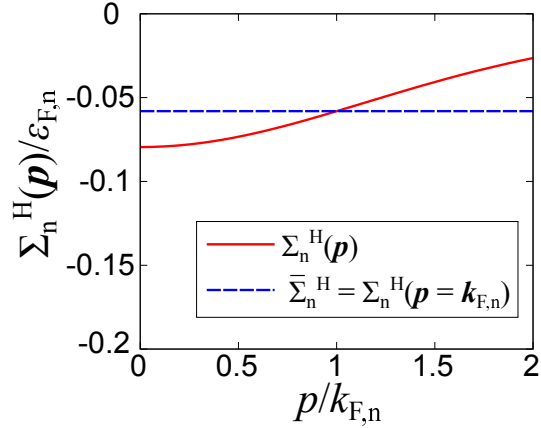}
\end{center}
\caption{Hartree self-energy $\Sigma_{\rm n}^{\rm H}(\bm{p})$ and the approximated Hartree shift $\bar{\Sigma}_{\rm n}^{\rm H}=\Sigma_{\rm n}^{\rm H}(\bm{p}=\bm{k}_{\rm F,n})$ in pure neutron matter with SEP at $\rho=0.29\rho_0$, $\varepsilon_{\rm F,n}=\mu_{\rm n}$ and $T=0.1\varepsilon_{\rm F,n}$}
\label{figa1}
\end{figure}
Figure \ref{figa1} shows the momentum dependence of the Hartree self-energy $\Sigma_i^{\rm H}(\bm{p})$ in the pure neutron matter at $\rho=0.29\rho_0$ with SEP1.  We set $\varepsilon_{\rm F,n}=\mu_{\rm n}$ and $T=0.1\varepsilon_{\rm F,n}$, and pairing-fluctuation effects are neglected for simplicity.
The magnitude of the Hartree shift is relatively small compared to the neutron chemical potential $\mu_{\rm n}$ and its
 momentum dependence is not substantial.
  Since 
 the momentum at the Fermi surface is the most important for Cooper pairings,
we introduce an approximation $\bar{\Sigma}_{i}^{\rm H}=\Sigma_{i}^{\rm H}(\bm{p}=k_{{\rm F},i})$ 
as adopted in the text.

In general, the momentum dependence of the Hartree self-energy near the Fermi surface gives rise to the 
effective mass $M^*$ defined by \cite{Jin}
\begin{eqnarray}
\label{eqA1}
\frac{1}{M^*}=\frac{1}{M}+\left.2\frac{\partial \Sigma_i^{\rm H}(\bm{p})}{\partial p^2}\right|_{\bm{p}=\bm{k}_{{\rm F},i}}.
\end{eqnarray}
From Fig. \ref{figa1}, we find  $M^*\simeq 0.98M$. In the present work, we have not taken into account this small correction.

We note that the present approximation of  the Hartree shift is different from the previous work~\cite{Pieter1}, where
$\bar{\Sigma}_{\rm n}^{\rm H}=V_{\rm s}(\bm{0},\bm{0})\rho_{\rm n}^{\rm H}/2=V_{\rm s}^{\rm SEP}(0,0)\rho_{\rm n}^{\rm H}/2$ is used.
While such an approximation of the  Hartree shift is sufficient enough  in the low-density region,
 it leads to the divergence of $L_{ij}$ near the nuclear saturation density.
  Furthermore, the present form of the  Hartree shift  is rather  
   consistent with the mean-field approximation under the separable interaction~$V_{\rm s}^{\rm SEP}(k,k')$.
   
\begin{figure}[t]
\begin{center}
\includegraphics[width=7cm]{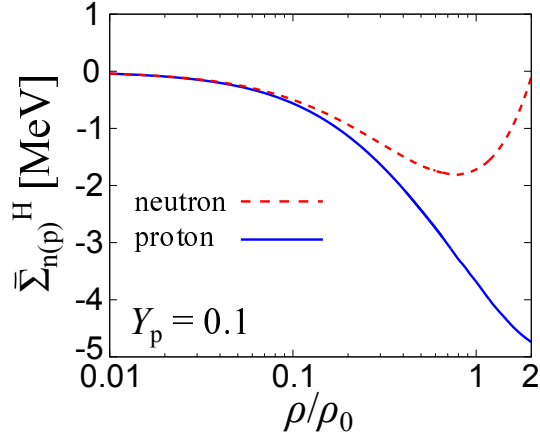}
\end{center}
\caption{Thouless determinant in the deuteron channel as a function of the center-of-mass momentum $q$ at $Y_{\rm p}=0.2$ for
 different values of the baryon density.}
\label{figa2}
\end{figure}
Figure \ref{figa2} shows the baryon density dependence of $\bar{\Sigma}_i^{\rm H}$ in asymmetric nuclear matter with $Y_{\rm p}=0.1$.
In the low-density limit, the shifts are negligibly small where the interaction can be well approximated by the contact-type interaction. 
While $\bar{\Sigma}_{\rm n}^{\rm H}$ increases around the nuclear matter density due to the short-range repulsion in the $^1S_0$ nn channel,
$\bar{\Sigma}_{\rm p}^{\rm H}$ decreases further, reflecting the difference between $k_{\rm F,p}$ and $k_{\rm F,n}$.
In addition, the behavior of $\bar{\Sigma}_{\rm p}^{\rm H}$ is mainly dictated by the $^3S_1$ np 
interaction  rather than the $^1S_0$ pp interaction because of $\rho_{\rm n}>\rho_{\rm p}$ in neutron star matter.

\section{$T_{\rm c}^{\rm pp}$ and $T_{\rm c}^{\rm d}$ at higher and lower densities}
\label{appB}
\begin{figure}[t]
\begin{center}
\includegraphics[width=7cm]{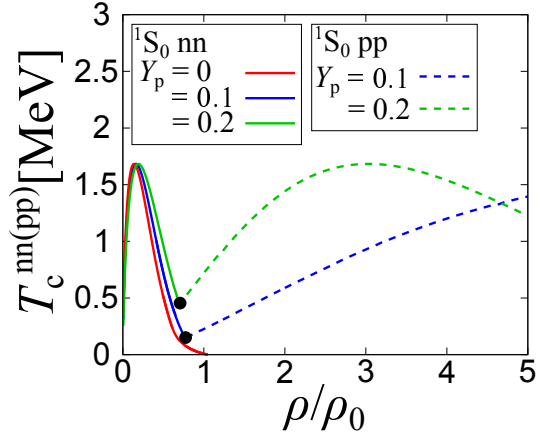}
\end{center}
\caption{Calculated critical temperature $T_{\rm c}^{\rm nn(pp)}$ of the $^1S_0$ neutron superfluidity (proton superconductivity) in asymmetric nuclear matter up to $\rho=5\rho_0$.}
\label{figb1}
\end{figure}
Since our separable interactions are adjusted so as to reproduce the AV18 phase shift up to $k=2$~ fm$^{-1}$, they cannot be used to investigate the properties of neutron matter above $\rho=1.59\rho_0$ (where $k_{\rm F,n}=2$~fm$^{-1}$).
On the other hand, the effective pp interaction $V_{\rm s}^{\rm SEP}(k_{\rm F,p},k_{\rm F,p})$ at the proton Fermi momentum $k_{\rm F,p}$ is still in the range of $0\leq k_{\rm F,p}\leq2$~fm$^{-1}$ even up to $\rho=15.9\rho_0$ 
in the case of $Y_{\rm p}=0.1$.  Therefore, just to see the qualitative behavior at high density,  we plot 
 $T_{\rm c}^{\rm pp}$ up to $5\rho_0$ in Fig .\ref{figb1}.  The result
  exhibits an upturn behavior in higher density regime due to the effective-range correction
   as well as short-range repulsion in the $^1S_0$ pp channel.
The $^3S_1$ np interaction modifies its density dependence through the suppression of the effective proton chemical potential $\mu_{\rm p}^{\rm H}$ as shown in Fig.~\ref{fig7} (b).
\par
\par
\begin{figure}[t]
\begin{center}
\includegraphics[width=7cm]{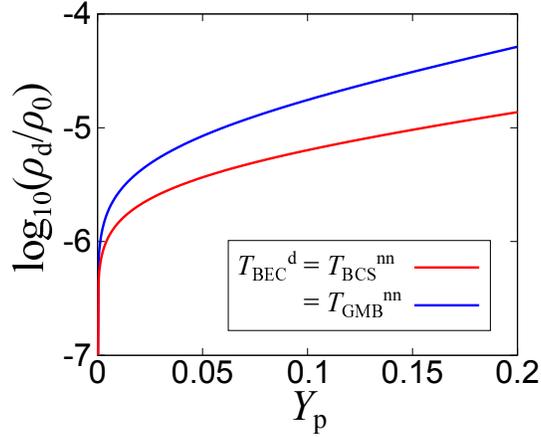}
\end{center}
\caption{The proton fraction dependence of the critical nucleon density $\rho_{\rm d}$ where $T_{\rm BEC}^{\rm d}=T_{\rm BCS}^{\rm nn}$ and $T_{\rm BEC}^{\rm d}=T_{\rm GMB}^{\rm nn}$.
}
\label{figb2}
\end{figure}
On the other hand, in the low-density limit, $T_{\rm BEC}^{\rm d}$ exceeds $T_{\rm c}^{\rm nn}$ in the case of a finite proton fraction.
In this limit, $T_{\rm c}^{\rm nn}$ is given by the zero-range BCS result
\begin{eqnarray}
\label{eqB1}
T_{\rm BCS}^{\rm nn}=\frac{8e^{\gamma}}{\pi e^2}\varepsilon_{\rm F,n}e^{\frac{\pi}{2k_{\rm F,n}a_{\rm s}}}.
\end{eqnarray}
where $\gamma=0.577$ is the Euler constant.
Since $T_{\rm c}^{\rm d}$ is equal to $T_{\rm BEC}^{\rm d}$ given by Eq. (\ref{eqdBEC}) due to the large binding energy $|E_{\rm d}|=2.22$ MeV,
we can analytically obtain the critical nucleon density $\rho_{\rm d}$ where $T_{\rm BEC}^{\rm d}=T_{\rm BCS}^{\rm nn}$ as
\begin{eqnarray}
\label{eqB2}
\rho_{\rm d}=\frac{\pi}{24a_{\rm s}^3(1-Y_{\rm p})}
\left[2\ln\left(\frac{\pi}{2}\right)+2-\gamma+\frac{2}{3}\ln\left(\frac{1}{9\pi^2\zeta(3/2)}\frac{Y_{\rm p}}{1-Y_{\rm p}}\right)\right]^{-3}.
\end{eqnarray}
Figure \ref{figb2} shows the proton fraction dependence of $\rho_{\rm d}$.
We also plot $\rho_{\rm d}$ obtained from the GMB result
$T_{\rm BEC}^{\rm d}=T_{\rm GMB}^{\rm nn}=(4e)^{-1/3}T_{\rm BCS}^{\rm nn}$ in the presence of the screening correction \cite{GMB}. 
In the relevant region for a neutron star ($0<Y_{\rm p}<0.2$),
$\rho_{\rm d}$ is smaller than the neutron drip density $\rho_{\rm drip}/\rho_0=1.5\times 10^{-3}$ \cite{Dean}.
We note that Eq. (\ref{eqB2}) is valid at small proton fraction ($Y_{\rm p}<0.2$), 
where $\rho_{\rm d}$ appears in the sufficiently low-density regime [$(k_{\rm F,n}a_{\rm s})^{-1}<-1$]~\cite{Tajima3}.


\begin{thebibliography}{99}


\bibitem{Takatsuka} T. Takatsuka and R. Tamagaki, Prog.  Theor. Phys. Suppl. \textbf{112}, 27 (1993).
\bibitem{Dean} D.-J. Dean and M. Hjorth-Jensen, Rev. Mod. Phys. \textbf{75}, 607 (2003).




\bibitem{Page3} D. Page, J. M. Lattimer, M. Prakash, and A. W. Steiner, in {\it Novel Superfluids}, edited by K. H. Bennemann and J. B. Ketterson (Oxford University Press, Oxford, 2014).
\bibitem{Oertel} M. Oertel, M. Henpel, T. Kl\"{a}hn, S. Typel, Rev. Mod. Phys. \textbf{89}, 015007 (2017).

\bibitem{Baym:2017whm} 
  G.~Baym, T.~Hatsuda, T.~Kojo, P.~D.~Powell, Y.~Song and T.~Takatsuka,
  Rept.\ Prog.\ Phys.\  {\bf 81}, 056902 (2018).

\bibitem{Carlson:2012mh} 
  J.~Carlson, S.~Gandolfi and A.~Gezerlis,
  PTEP {\bf 2012}, 01A209 (2012).
%


\bibitem{Gandolfi} S. Gandolfi, A. Gezerlis, and J. Carlson,
Annu. Rev. Nucl. Part. Sci. \textbf{65}, 303 (2015).
\bibitem{Horikoshi} M. Horikoshi, M. Koashi, H. Tajima, Y. Ohashi, and M. Kuwata-Gonokami, Phys. Rev. X \textbf{7}, 041004 (2017). 
\bibitem{HorikoshiE} M. Horikoshi and M. Kuwata-Gonokami, Int. J Mod. Phys. E \textbf{28}, 1930001 (2019).
\bibitem{Strinati} G. C. Strinati, P. Pieri, G. R\"{o}pke, P. Schuck, and M. Urban, Phys. Rep. \textbf{738}, 1 (2018).

\bibitem{Chin} C. Chin, R. Grimm, P. Julienne, and E. Tiesinga, Rev. Mod. Phys. \textbf{82}, 1225 (2010).


\bibitem{Mueller} E. J. Mueller, Rep. Prog. Phys. \textbf{80}, 104401 (2017).
\bibitem{Jensen} S. Jensen, C. N. Gilbreth, and Y. Alhassid, Eur. Phys. J. Spec. Top. \textbf{227}, 2241 (2019).


\bibitem{Pieter1} P. van Wyk, H. Tajima, D. Inotani, A. Ohnishi, and Y. Ohashi, Phys. Rev. A \textbf{97}, 013601 (2018).



\bibitem{Nozieres} P. Nozi\`{e}res and S. Schmitt-Rink, J. Low Temp. Phys. \textbf{59}, 195 (1985).

\bibitem{Yamaguchi} Y. Yamaguchi, Phys. Rev. \textbf{95}, 1628 (1954).
\bibitem{Mongan} T. R. Mongan, Phys. Rev. \textbf{175}, 1260 (1968).
\bibitem{Mongan2} T. R. Mongan, Phys. Rev. \textbf{178}, 1597 (1969).
\bibitem{Mathelitsch} L. Mathelitsch, W. Plessas, and W. Schweiger, Phys. Rev. C \textbf{26}, 65 (1982). 
\bibitem{Haidenbauer} J. Haidenbauer and W. Plessas, Phys. Rev. C \textbf{30}, 1822 (1984).
\bibitem{Haidenbauer3} J. Haidenbauer and W. Plessas, Phys. Rev. C \textbf{32}, 1424 (1985).
\bibitem{Grygorov} P. Grygorov, E. N. E. van Dalen, H. M\"{u}ther, and J. Margueron, Phys. Rev. C \textbf{82}, 014315 (2010).

\bibitem{Alm} T. Alm, B. L. Friman, G. R\"{o}pke, and H. Schulz, Nucl. Phys. A \textbf{A551}, 45 (1993).
\bibitem{Osman} A. Osman, Phys. Rev. C \textbf{19}, 1127 (1979).
\bibitem{Schnell} A. Schnell, G. R\"{o}pke, and P. Schuck, Phys. Rev. Lett. \textbf{83}, 1926 (1993).
\bibitem{Sedrakian2} A. D. Sedrakian, D. Blaschke, G. R\"{o}pke, and H. Schulz, 
Phys. Lett. B \textbf{338}, 111 (1994).
\bibitem{Beyer} M. Beyer, G. R\"{o}pke, and A. Sedrakian, 
Phys. Lett. B \textbf{376}, 7 (1996).
\bibitem{Schadow} W. Schadow, W. Sandhas, J. Haidenbauer, and A. Nogga,
Few-Body Syst.\textbf{28}, 241 (2000).
\bibitem{Bozek} P. Bo\'{z}ek and P. Czerski, Eur. Phys. J. A \textbf{qq}, 271 (2001).
\bibitem{Dewulf} Y. Dewulf, W. H. Dickhoff, D. Van Neck, E. R. Stoddard, and M. Waroquier, Phys. Rev. Lett. \textbf{90}, 152501 (2003).
\bibitem{Stein} H. Stein, A. Schnell, T. Alm and G. R\"{o}pke, Z. Phys. A {\bf 351}, 295 (1995)
\bibitem{Jin} M. Jin, M. Urban, and P. Schuck, Phys. Rev. C \textbf{82}, 024911 (2010).
\bibitem{Martin} N. Martin and M. Urban, Phys. Rev. C \textbf{90}, 065805 (2014).

\bibitem{AV18} R. B. Wiringa, V. G. J. Stoks, and R. Schiavilla, Phys. Rev. C \textbf{51}, 38 (1995).


\bibitem{Thouless} D. J. Thouless, Annals of Physics \textbf{10}, 553 (1960).

%
%
%
%
%
%
%
%














\bibitem{Ramanan} S. Ramanan and M. Urban, Phys. Rev. C \textbf{88}, 054315 (2013).







\bibitem{note} We note that since the calculation of $T_{\rm c}$ within SEP3 in the high density region $k_{\rm F,n}\gesim 1.3$ fm$^{-1}$ of pure neutron matter is numerically demanding,
we extrapolate them to $k_{\rm F,n}=1.73$ fm$^{-1}$ where $T_{\rm c}$ invariably disappears because the phase shift at $k=k_{\rm F,n}$ becomes zero there, by using the Pad\'{e} approximation.

\bibitem{Abe} T. Abe and R. Seki, Phys. Rev. C \textbf{79}, 054003 (2009).


\bibitem{Andrenacci} N. Andrenacci, A. Perali, P. Pieri, and G. C. Strinati, Phys. Rev. B \textbf{60}, 12 410 (1999).

\bibitem{Tajima3} H. Tajima, T. Hatsuda and Y. Ohashi, J. Phys.: Conf. Ser. \textbf{969}, 012003 (2018).


\bibitem{Lattimer} J. M. Lattimer, C. J. Pethick, M. Prakash, and P. Haensel, Phys. Rev. Lett. \textbf{66}, 2701 (1991).
\bibitem{APR} A. Akmal, V. Pandharipande, and D. Ravenhall, Phys. Rev. C \textbf{58}, 1804 (1998).
\bibitem{Alford} M. G. Alford and S. P. Harris, Phys. Rev. C \textbf{98}, 065806 (2018).


\bibitem{Liu} X.-J. Liu and H. Hu, Europhys. Lett. \textbf{75}, 364 (2006).

\bibitem{FF} P. Fulde and R. A. Ferrell, Phys. Rev. \textbf{135}, A550 (1964).
\bibitem{LO} A. I. Larkin and Y. N. Ovchinnikov, Sov. Phys. JETP \textbf{20}, 762 (1965).
\bibitem{Sheehy} L. Radzihovsky and D. E. Sheehy, Rep. Prog. Phys. \textbf{73}, 076501 (2010).
\bibitem{Radzihovsky} L. Radzihovsky, Phys. Rev. A \textbf{84}, 023611 (2011).
\bibitem{OhashiFF} Y. Ohashi, J. Phys. Soc. Jpn. \textbf{71}, 2625 (2002).
\bibitem{FrankZwerger} B. Frank, J. Lang, and W. Zwerger, J. Exp. Theor. Phys. \textbf{127}, 812 (2018).

Phys. Rev. C \textbf{64}, 064314 (2001).


\bibitem{GMB} L. P. Gorkov and T.  K. Melik-Barkhudarov, Sov. Phys. JETP \textbf{13}, 1018 (1961) [Zh. Eksp. Teor. Fiz. \textbf{40}, 1452 (1961)].

\bibitem{Yu} Z.-Q. Yu, K. Huang, and L. Yin, Phys. Rev. A \textbf{79}, 053636 (2009).
\bibitem{Pisani} L. Pisani, A. Perali, P. Pieri, and G. C. Strinati, Phys. Rev. B \textbf{97}, 014528 (2018).

\bibitem{Cao} L. G. Cao, U. Lombardo, and P. Schuck, Phys. Rev. C \textbf{74}, 064301 (2006).


\bibitem{Ramanan2} S. Ramanan and M. Urban, Phys. Rev. C \textbf{98}, 024314 (2018).

\bibitem{Tamagaki} R. Tamagaki, Prog. Theor. Phys. \textbf{44}, 905 (1970).
\bibitem{Hoffberg} M. Hoffberg, A. E. Glassgold, R. W. Richardson, and M. Ruderman, Phys. Rev. Lett. \textbf{24}, 775 (1970).
\bibitem{Takatsuka2} T. Takatsuka, Prog. Theor. Phys. . \textbf{48}, 1517 (1972).


























 









%


\end{thebibliography}

\end{document}